# MEASURING THE EVOLUTION OF THE M/L RATIO FROM THE FUNDAMENTAL PLANE IN CL 0024+16 AT $Z=0.39$

MARIJN FRANX and PIETER G. VAN DOKKUM

*Kapteyn Astronomical Institute*
*P.O. Box 800, 9700 AV Groningen*
*The Netherlands*

**Abstract.** The existence of the Fundamental Plane of early-type galaxies implies that the $M/L$ ratios of early-types are well behaved. It provides therefore an important tool to measure the evolution of the $M/L$ ratio with redshift. These measurements, in combination with measurements of the evolution of the luminosity function, can be used to constrain the mass evolution of galaxies.

We present the Fundamental Plane relation measured for galaxies in the rich cluster CL 0024+16 at $z$=0.391. The galaxies satisfy a tight Fundamental Plane, with relatively low scatter (15 %). The $M/L$ is 31 ± 12 % lower than the $M/L$ measured in Coma, which is consistent with simple evolutionary models. Hence, galaxies with very similar dynamical properties existed at a $z$=0.4.

More, and deeper data are needed to measure the evolution of the slope and the scatter of the Fundamental Plane to higher accuracy. Furthermore, data on the richest nearby clusters would be valuable to test the hypothesis that the Fundamental Plane is independent of cluster environment.

## 1. Introduction

Galaxy evolution may be a complex process, with possibly a large role for mergers, interactions, infall, and starbursts triggered by these events. Such processes complicate the interpretation of observations of high redshift galaxies, as galaxies can change rapidly in luminosity (due to starbursts), and can change morphology due to mergers, and infall of gas. The progen-itors of certain types of galaxies at some redshift may be of different type at some other redshift, and their luminosities may be quite different.

In order to quantify these effects, more information is needed than the evolution of luminosity and color of galaxies. Detailed information on the morphological evolution, and the evolution of the mass function is essential. The evolution of the mass function is possibly the most important, as it gives direct insight into the mass evolution of individual galaxies, and can directly determine when typical galaxies were assembled.

Unfortunately, the total masses of galaxies are notoriously difficult to measure. However, there exist good relations between circular velocity, and velocity dispersion, and photometric parameters: the Tully-Fisher relation for spirals (Tully & Fisher 1977), the Faber-Jackson relation (Faber & Jackson 1976), and the Fundamental Plane for early-types (Djorgovski & Davis 1987, Dressler et al. 1987). These relations are very suitable for evolutionary studies, because their intrinsic scatter is low at $z=0$.

Here we present results on a program to measure the evolution of the Fundamental Plane relation with redshift. The Fundamental Plane is a relation between effective radius $r_e$, effective surface brightness $I_e$, and central velocity dispersion $\sigma$ of the form $r_e \propto \sigma^{1.24} I_e^{-0.82}$ (e.g., Bender et al. 1992, Jørgensen et al 1995, JFK). Its scatter is low, at 17% in $r_e$ (Lucey et al. 1991, JFK). The implication of the Fundamental Plane is that the $M/L$ ratio of galaxies is well behaved (e.g., Faber et al. 1987). Under the assumption that galaxies are a homologous family, the implied $M/L$ scaling is $M/L \propto r_e^{0.22} \sigma^{0.49} \propto M^{0.24}$. Such scaling is sufficient for the existence of the Fundamental Plane, and vice versa. The cause of the variation in $M/L$ with mass parameters is not well understood, but it can be due to variations in metallicity, IMF, dark matter fraction, and age (e.g., Renzini & Ciotti 1993).

## 2. What can we hope to learn ?

The low intrinsic scatter of the Fundamental Plane might make it suitable for "classical" cosmological tests. These tests would require that the evolutionary effects can be ignored. The following applications can be considered:

• Surface brightness test ? The Fundamental Plane can be used to perform the classical surface brightness test (Kjærgaard et al. 1993). However, this surface brightness test has done by COBE. If surface brightness were not to decrease with $(1+z)^{-4}$, the Cosmic Background Radiation would deviate from a Planck spectrum on a very short time-scale. Since it satisfies a Planck spectrum to 0.03 % (Mather et al. 1994), the cosmological dimming holds to very high accuracy.

• $q_0$ - Apparent angle of a standard rod ? We can define as a standard rod

the observed effective radius of some galaxy with fixed surface brightness, and velocity dispersion. The dependence against redshift might provide a constraint on $q_0$, except that evolutionary effects in the luminosity of galaxies are larger than this cosmological effect.

These cosmological tests are therefore of limited use. The most important application of the evolution of the Fundamental Plane is the measurement of the evolution of the $M/L$ ratio. This measurement can be used to determine the evolution of the mass function, if the evolution of the luminosity function is also well determined. Furthermore, the scatter and the slope of the Fundamental Plane may change with redshift. The measurement of this evolution will provide additional insight into the cause of the dependence of the $M/L$ ratio on mass (Renzini & Ciotti 1993).

## 3. Models for the evolution of the $M/L$ ratio

The luminosity of a co-eval stellar population is expected to evolve with time. This is due to the fact that much of the light is produced by stars on the giant branch, which is a very short phase in a stellar lifetime. Tinsley (1980) showed that the luminosity evolves like

$$L \propto 1/(t - t_{form})^\kappa$$

where $\kappa = 1.3 - 0.3x$, and $x$ is the slope of the IMF. The Miller–Scalo IMF implies $x=0.25$, and $\kappa \approx 1.2$. Recent studies indicate that the value of $\kappa$ depend on passband and metallicity (Buzzoni 1989, Worthey 1994). These authors find $0.6 < \kappa < 0.95$ for the V band.

To first order, this evolution implies that the $M/L$ ratio evolves like

$$\ln M/L(z) = \ln M/L(0) - \kappa(1 + q_0 + 1/z_{form})\,z,$$

where $z_{form}$ is the formation redshift (Franx 1995). Hence the logarithm of the $M/L$ ratio is expected to decrease linearly with redshift, and the coefficient depends on $\kappa$(IMF), $q_0$, and $z_{form}$. This equation is valid for $q_0 \approx 0$, and high $z_{form}$, but it is a reasonable approximation even for $q_0 = -1$. The rate at which the $M/L$ ratio decreases is therefore a function of several unknown variables, and the interpretation of the observed decrease of the $M/L$ ratio will not be very straightforward.

### 3.1. COMPLEX EVOLUTION

There is no good reason to assume that all early-type galaxies formed at the same redshift. Furthermore, a single galaxy may have had a complex formation history, with starformation extending over a long time. The evolution of the $M/L$ ratio will be much more complex if such age differences are taken into account.

If galaxies have different mean ages, then the $M/L$ ratios will be different (if other properties are the same). Hence, there will be scatter in the $M/L$ ratios, and the scatter will increase with look-back time, as the relative age difference will increase. Such effects can therefore be found by measuring the scatter in the Fundamental Plane as a function of redshift.

Galaxy evolution can be substantially more complex than that, however, as demonstrated by the Butcher-Oemler effect (Butcher & Oemler 1978, 1984), and the presence of many post starburst galaxies in clusters (Dressler & Gunn 1983, Couch & Sharples 1987). We have constructed simple models to evaluate the effect of such evolution on the observed $M/L$ ratios. We assume that galaxies are assembled at $z=4$, and form stars in a very regular way. This occurs presumably in a disk. Then, at some time, they undergo a strong starburst, in which 10% of their mass is converted into stars, and they cease forming stars. Their spectra will be characteristic of post starburst galaxies for another 1.5 Gyr, and after that they will classified as normal early-type galaxies. This type of evolution implies that the morphologies of galaxies evolve with time, from spiral, to post star burst galaxy, to early-type. This has important consequences, since the set of early-types at higher redshifts will be a special subset of the set of early-types at $z=0$. If we select early-types at higher and higher redshift, we are selecting a subsample that is more and more biased towards the oldest early-types. In short, we may be selecting the oldest galaxies, and find that they are old.

The problem is illustrated in Fig. 1. The typical evolution of 3 galaxies is shown. The thick line is the phase in which they appear as early-types. Clearly, the oldest early-types appear as early-type for the longest time. Fig. 1b demonstrates the effect on the observed $L/M$ ratios of a large sample. At low redshifts, all galaxies appear as early types, and the evolution of the median $L/M$ ratio remains normal. The scatter around the mean increases rapidly with redshift. Around $z = 0.2$, some of the galaxies appear as post star burst galaxies, and they would be excluded. The median $L/M$ ratio is biased towards low values. This effects increases at higher redshifts. The bias is as strong as 30 % at $z=0.5$. As galaxies disappear from the sample, the scatter in the $L/M$ ratio may decrease at higher redshifts.

Many variations of such models can be made, and they will produce similar effects. These models have one common aspect: they predict that the morphologies of galaxies evolve with redshift. The morphologies of galaxies in distant clusters can now be determined accurately from HST images (e.g., Dressler et al. 1994). In combination with good spectral information, the morphological evolution can be measured directly.

The effect of this complex evolution will be similar on other properties such as colors, and linestrengths. The scatter is expected to increase at lower redshifts, and may actually decrease at higher redshifts. The median color evolution may show a "break", at the redshift where galaxies start disappearing from the sample.

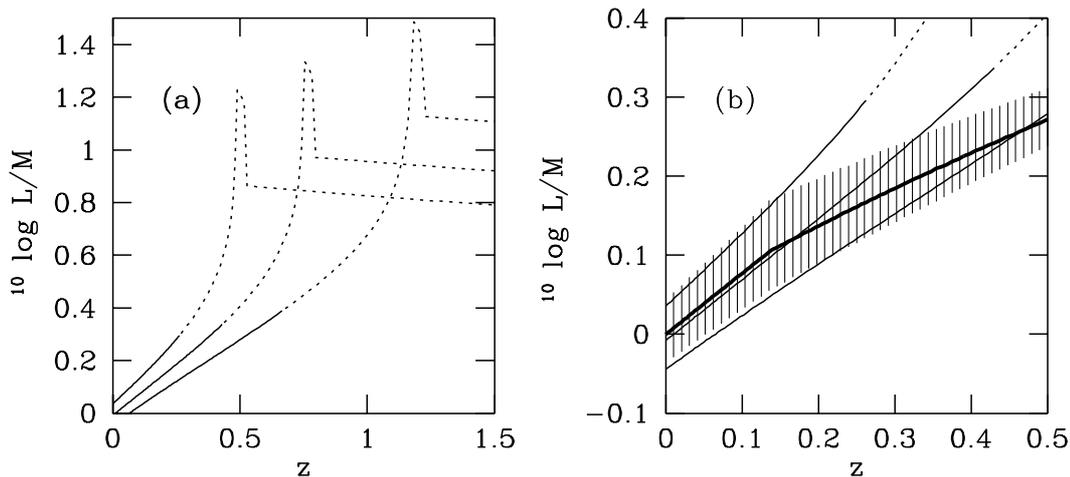

*Figure 1.* The evolution of galaxies which undergo three distinct phases: I regular star formation in a disk, II starburst, III quiescent evolution. a) shows the luminosity evolution for three such galaxies. The galaxies are classified as regular early-types after 1.5 Gyr after the burst. This is indicated by the thick line. b) the evolution of the mean $L/M$ ratio for a sample of early-types which formed in this complex way. The starburst is assumed to occur at a random time between $z=0.5$ and $z=2$. The line indicates the median $L/M$, the shaded area is bounded by the upper and lower quartile of the sample. The median $L/M$ ratio bends at $z=0.2$, as more and more galaxies drop out from the sample. The sample becomes more and more biased to the oldest early-types at higher redshifts.

## 4. The Fundamental Plane in CL 0024+16 at $z=0.39$

CL 0024 is a rich cluster at $z=0.39$, and has been extensively observed (e.g., Dressler et al. 1985). We have obtained a deep, 19 hour integration at the MMT to measure the internal velocity dispersions of luminous galaxies in the cluster. In total, 13 galaxies were observed, and velocity dispersions were measured for 9. The dispersions were measured in the same way as for low redshift galaxies, with the important difference that the instrumental resolution had to be determined in an absolute way. This is due to the fact that it is impossible to obtain good template star spectra with the same instrumental setup (Franx, 1993a,b). Full details of the observations and the analysis can be found in van Dokkum and Franx (1996).

HST images were used to measure the length scales of the galaxies. The procedure consisted of convolving an $r^{1/4}$ model with the Point Spread Function, and fitting it directly to the data. In this fashion, the parameters $r_e$ and $I_e$ could be determined reliably. Although the errors in $r_e$ and $I_e$ can be substantial, the error in $r_e I_e^{0.8}$ is very small (on the order of 5%). The R photometry was reduced to restframe V band photometry, which made the color corrections very small.

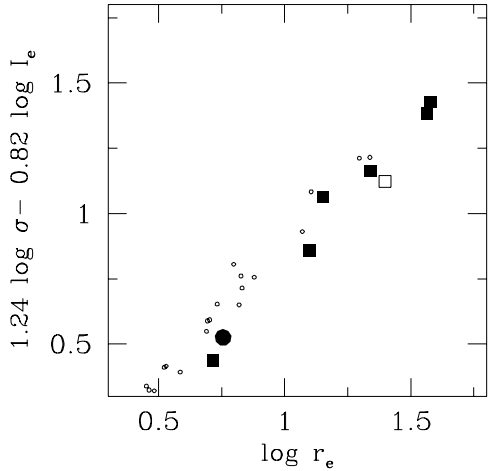

*Figure 2.* The Fundamental Plane for galaxies in CL 0024+16 at $z=0.391$ in the V band. The small symbols are galaxies in Coma. The Fundamental Plane in CL 0024 is very similar to that in Coma, with similar low scatter (15%).

Fig. 2 shows the resulting Fundamental Plane. There is a very clear relation, with relatively low scatter (15 %). The slope is very similar to that for nearby cluster galaxies (e.g., JFK). In short, *early-type galaxies exist at z=0.4 which are very similar to galaxies at z=0*. This extends earlier work on the colors, and color-magnitude relation of early-type galaxies (e.g., Bower et al. 1992).

The evolution of the $M/L$ ratio can be determined from the data. Fig. 3a shows the observed $M/L$ ratios for Coma and CL 0024 against the parameter $r_e^{0.22}\sigma^{0.49}$. The Fundamental Plane implies that galaxies lie along a line in the plot. If galaxies undergo only passive evolution, then the parameters $r_e$ and $\sigma$ remain constant with time. The evolution will cause a vertical shift with redshift. We see a clear offset between the two data sets. The lines indicate fits to both data sets. The mean difference in the $M/L$ ratio is 31 %. The error is dominated by systematic effects, and is estimated at 12 %. It is clear that the sample for CL0024 is biased towards the most massive galaxies, and this selection bias is partly the cause for the systematic uncertainty.

Fig. 3b shows the evolution of the $M/L$ ratio with redshift. The current results are in good agreement with a large formation redshift, and small values of $\kappa$. The resulting constraint is $\kappa(1 + q_0 + 1/z_{form}) = 0.84 \pm 0.32$. Obviously, many combinations of $q_0$ and $z_{form}$ are allowed, given the uncertainty in $\kappa$ and the observed change in $M/L$ ratio.

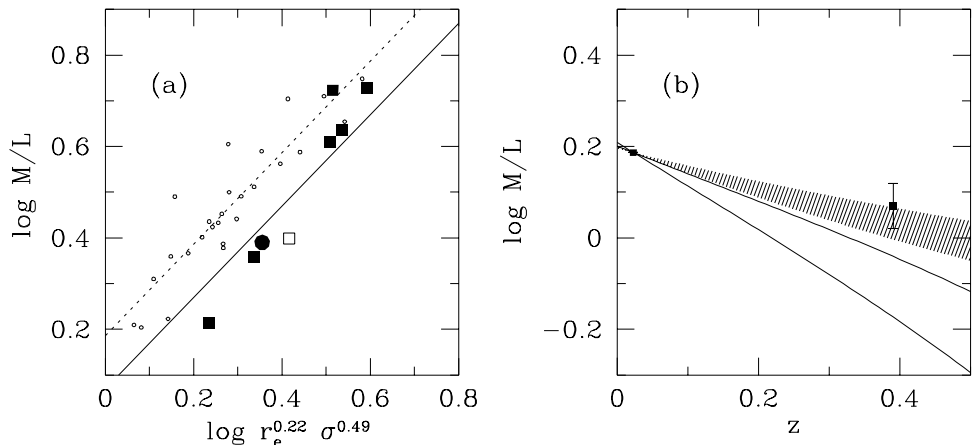

*Figure 3.* a) The $M/L$ ratio against $r_e^{0.22}\sigma^{0.49} \propto M^{0.24}$, for CL 0024 and Coma. The lines are fits to the data points. The $M/L$ ratio in CL 0024 is lower by 31± 12 %. b) The evolution of the $M/L$ ratio against redshift. The two points are Coma and CL 0024. The hatched area is the area allowed by models with formation redshift $z_{form} = \infty$. The two lines delineate the area allowed by models with $z_{form} = 1$. Such models are marginally consistent with the data.

## 5. Discussion

We have shown that the equivalent of the Fundamental Plane exists at $z=0.39$, and we have derived the mean evolution of the $M/L$ ratio. The evolution is low at 31%, but is in agreement with models. The uncertainties in the modeling still allow a wide range in formation redshifts $z_{form}$.

The next step is to observe more and fainter galaxies in distant clusters, and to observe more clusters. Such measurements can address questions about the change in slope of the Fundamental Plane, and the dependence of galaxy evolution on environment.

The results can be compared in a qualitative way to measurements of the evolution of the luminosity function in clusters and the field. The evolution of the luminosity function gives additional information on the luminosity evolution of individual galaxies, with the added uncertainty that dissipationless merging and infall can affect the luminosities greatly. The $M/L$ ratio is much less sensitive to such processes. Aragon-Salamanca et al. (1993) found that the K luminosity function of cluster galaxies does not evolve with redshift. The uncertainties are still large, however, and the measurements are not inconsistent with the evolution of the $M/L$ as measured from the Fundamental Plane. Lilly et al. (1995) and Ellis et al. (1996) found very weak evolution of the bright end of the luminosity function in the field. Lilly et al. took one further step, and derived very weak evolution for red galaxies in general. All these measurements are qualitatively in agreement with the current results. As more data will be gathered on the evolution of the luminosity function and the $M/L$ ratio, the general mass evolution of galaxies may be determined from such data in the future.

Studies of color evolution of galaxies in clusters indicate weak evolution for $0 < z < 0.5$ (Aragon-Salamanca et al. 1993, Rakos & Schombert 1995). Since stellar population models predict rapid color evolution when the luminosity evolution is slow, it is not clear yet how good these measurements can be reproduced by the models.

It is a pleasure to thank the organizers for a stimulating conference, and for financial support.